\pdfoutput=1
\documentclass{PoS}
\usepackage{graphics}
\usepackage{amsmath}

\title{A Fitting Robot for Variational Analysis}

\ShortTitle{A Fitting Robot}

\author{\speaker{Alan \'O Cais}\thanks{We thank the Australian Partnership for Advanced Computing (APAC) and the South Australian Partnership for
Advanced Computing (SAPAC) for generous grants of supercomputer time which have enabled this project. We also thank the University of Adelaide Faculty of Sciences for allowing us time on their Condor\texttrademark cycle-scavenging system. This work is supported by the Australian Research Council.}\\
        University of Adelaide, Adelaide, SA 5005, Australia\\
        E-mail: \email{alan.ocais@cyi.ac.cy}}

\author{Derek Leinweber\\
        University of Adelaide\\
        E-mail: \email{derek.leinweber@adelaide.edu.au}}

\author{Selim Mahbub\\
        University of Adelaide\\
        E-mail: \email{md.mahbub@adelaide.edu.au}}

\author{Tony Williams\\
        University of Adelaide\\
        E-mail: \email{anthony.williams@adelaide.edu.au}}

\abstract{We develop a robot algorithm to maximise the number of distinct states reliably extracted from correlator data using the variational analysis method. The robot explores the variational parameter space and attempts to remove, as far as possible, the human element from the fitting of the subsequent orthogonalised data.}

\FullConference{The XXVI International Symposium on Lattice Field Theory\\
		 July 14-19 2008\\
		 Williamsburg, Virginia, USA}

\begin{document}

\section{Introduction}

The variational approach \cite{Luscher:1990ck} is a widely used method to extract the excited state spectrum from matrices of correlation functions. Apart from the choice of operator basis, the variational method has a parameter space that should be expected to yield consistent results for a range of values (within certain limitations \cite{Blossier:2008tx}). Exploring this parameter space potentially yields large number of correlation functions to fit. 

We also currently have access to accurate methods for the determination of all-to-all quark propagators, e.g. Refs~\cite{Wilcox:1999ab, Neff:2001zr,  Foley:2005ac, Collins:2007mh, Boyle:2008rh}, which allows easy access to a very large number of available basis states for the variational approach. This in turn leads to another multiplicative factor of correlation functions to be fitted. Ultimately we must automate the procedure for fitting these data sets since, at some point, it becomes unfeasible to do it by hand. 

Here we describe both these processes and formulate a fitting procedure that attempts to combine a fitting algorithm and an exploration of the variational parameter space. It strives to deliver figures from correlation matrix analysis which can be used to determine consistent and robust fitted energies for multiple energy levels.

\subsection{Variational Analysis}
Extracting excited-states using a variational approach requires a matrix of correlation functions 
\begin{displaymath}
C_{\alpha\beta} (t ) = \langle 0| O_\alpha (t )O_\beta^\dagger (0)| 0\rangle
\end{displaymath} 
where $O_\alpha$ ($\alpha = 1\dots N$) form the $N$ operators with the appropriate quantum numbers of our variational basis. We define the $N$ principal correlators $\lambda_\alpha (t , t_0 )$ as the eigenvalues of $C (t_0 )^{ −1/ 2} C (t ) C (t_0 )^{−1/ 2}$ (where $t_0$ is the time defining the ``metric''). $t_0$ is (in theory) sufficiently large that one believes that the correlator at $t_0$ is dominated by the lightest $N$ states and (in practice) sufficiently small that the inversion of $C(t_0)$ does not become numerically unstable. We can then show \cite{Luscher:1990ck} that 
\begin{displaymath}
\displaystyle \lim_{t \to \infty } \lambda_\alpha (t , t_0 ) = e^{-( t - t_0 ) E_\alpha}(1 + \textrm{O}(e^{-t\Delta E_\alpha}) ) \, \textrm{ \, \, \, where \, } \, \Delta E_\alpha=\min_{\beta \neq \alpha}|E_\beta - E_\alpha| \, .
\end{displaymath}
                                                         
The $N$ principal effective masses defined by $m_\alpha^{eff}(t) = ln(\frac{\lambda_\alpha (t , t_0 )}{\lambda_\alpha (t+1 , t_0 )})$ now tend (plateau) to the $N$ lowest-lying stationary-state energies. Since all the time dependence of the diagonalisation process is contained in the principal correlators we can perform the diagonalisation at a single ``optimising'' time, $t=t_{opt}$, and the eigenvectors determined at this timeslice can be used to diagonalise the correlation matrix at all times. We can also then extract the effective masses by fitting the plateau regions of the $N$ diagonal entries of the projected correlation matrix.

This method relies on constructing a basis of operators $O_\alpha$ that provides a good description of the states of interest. With the advent of all-to-all propagator techniques we can easily construct a fuller variational basis from a combination of the multiple operators for each $O_h$ representation state (for example) and from the application of different smearing operators to the quark fields.

\subsection{All-to-all Propagators}
Many algorithms have been devised to approximate all-to-all propagators, for example see Refs. \cite{Wilcox:1999ab, Neff:2001zr}. Here we use an exact
implementation involving a hybrid method that combines an eigenvector
decomposition with a variance-reduced stochastic estimator \cite{Foley:2005ac}.

To construct the all-to-all propagator, the lowest $N_{\rm ev}$ eigenmodes (eigenvectors $v^{(i)}$ with eigenvalues $\lambda_i$) of
the hermitian Dirac matrix $Q = \gamma_5 M$ are first computed, and a truncated
spectral decomposition of the propagator is then given by
\begin{equation}
\bar{Q}_0=\sum_{i}^{N_{\rm ev}}\frac{1}{\lambda_i}v^{(i)}\otimes
 v^{(i)^\dag}\,.
\end{equation}
If $N_{\rm ev}$ is equal to the dimension of the matrix, then 
$\bar{Q}_0=Q^{-1}$, otherwise the propagator can be expressed as 
\begin{equation}
Q^{-1} = \bar{Q}_0 + \bar{Q}_1\, ,
\end{equation}
and the truncation in the eigenvector representation can be corrected by
estimating $\bar{Q}_1$ stochastically.

We choose to ``dilute'' the noise vectors, $\eta^{(i)}$, used in the stochastic estimation, which results in rapid variance 
reduction. In this context, dilution means creating a set of noise vectors
by applying a set of masks to a single noise source. These masks might for 
example select a particular time-slice of the vector (referred to as time 
dilution), thus returning $N_{\rm T}$ noise vectors from each single noise source.

The method can be implemented efficiently in software by use of a 
``hybrid list'' method. Two lists of $N_{\rm HL}=N_{\rm ev} + N_{\rm dil}$ vectors $u$ and 
$w$ are written, 
\begin{equation}
w^{(i)}=\biggl\{\frac{v^{(1)}}{\lambda_1},\cdots,\frac{v^{(N_{\rm ev})}}{\lambda_{N_{\rm ev}}},\eta^{(1)},\cdots,\eta^{(N_{\rm dil})}\biggr\}, \textrm{ \, \, \,}
u^{(i)}=\biggl\{v^{(1)},\cdots,v^{(N_{\rm ev})},\psi^{(1)},\cdots,\psi^{(N_{\rm dil})}\biggr\}\, ,
\end{equation}
where $M\psi^{(i)} = \gamma_5 (I - P_0)\eta^{(i)}$ and $P_0 = \sum_{i}^{N_{\rm ev}}v^{(i)}\otimes v^{(i)^\dag}$. The inverse of $M$ can then be written as a single sum over the pair of lists, $M^{-1}=\sum_{i=1}^{N_{\rm HL}}( u^{(i)}\otimes w^{(i)\dagger})\gamma_5\, .$ In all data contained in this paper 20 eigenmodes and dilution in time and colour space is used. 

\begin{figure*}[ht]
$\begin{array}{c@{\hspace{1cm}}c}
  \includegraphics[height=3.6cm,width= 6.9cm]{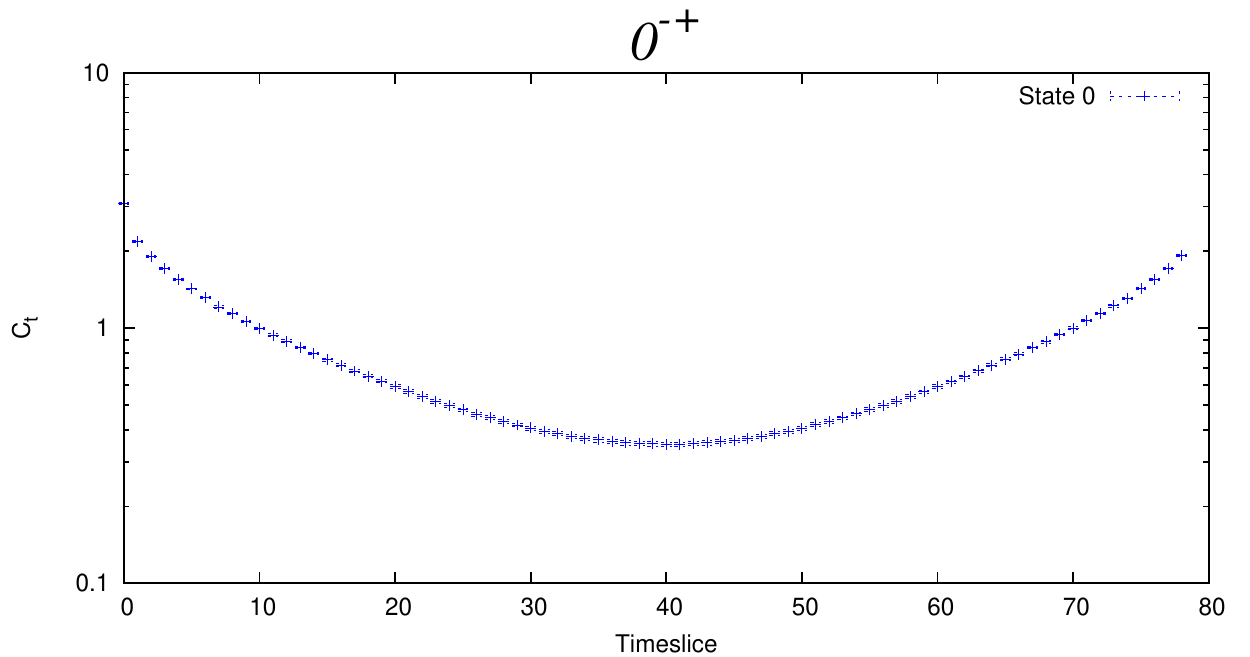} &
  \includegraphics[height=3.6cm,width= 6.9cm]{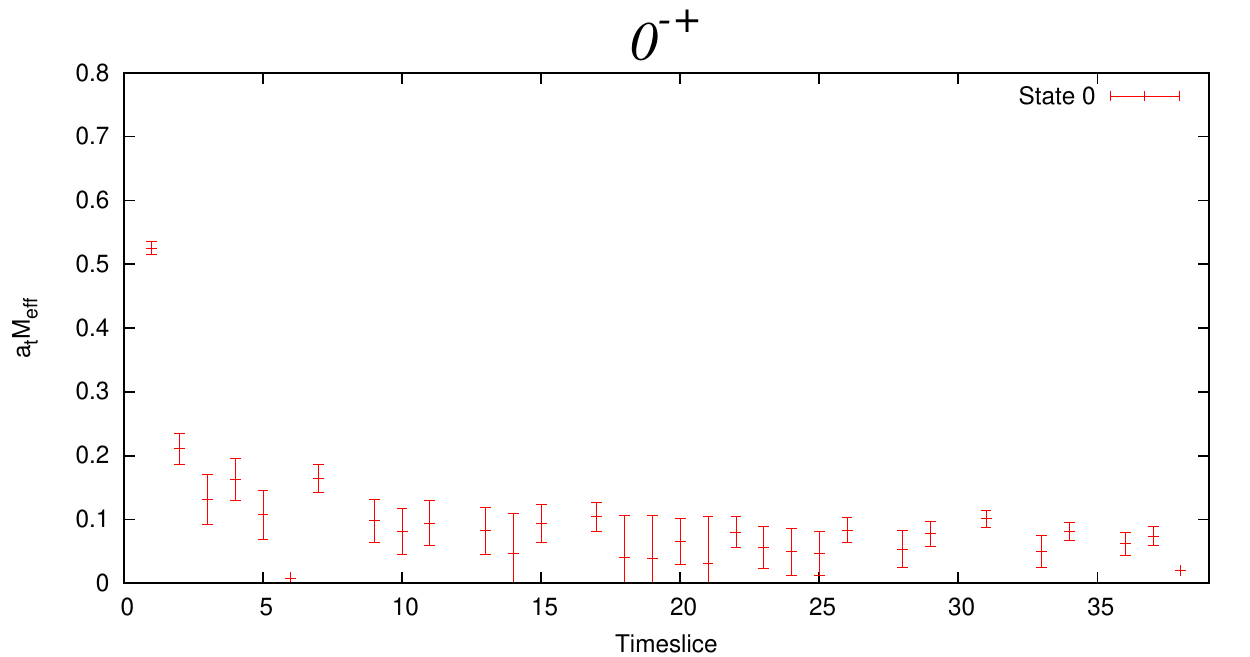}
  
\end{array}$
\caption{Correlation (left) and effective mass (right) plots for the pion ground state.}
\label{0mp}
\end{figure*}

\begin{figure*}[ht]
$\begin{array}{c@{\hspace{1cm}}c}
  \includegraphics[height=3.6cm,width= 6.9cm]{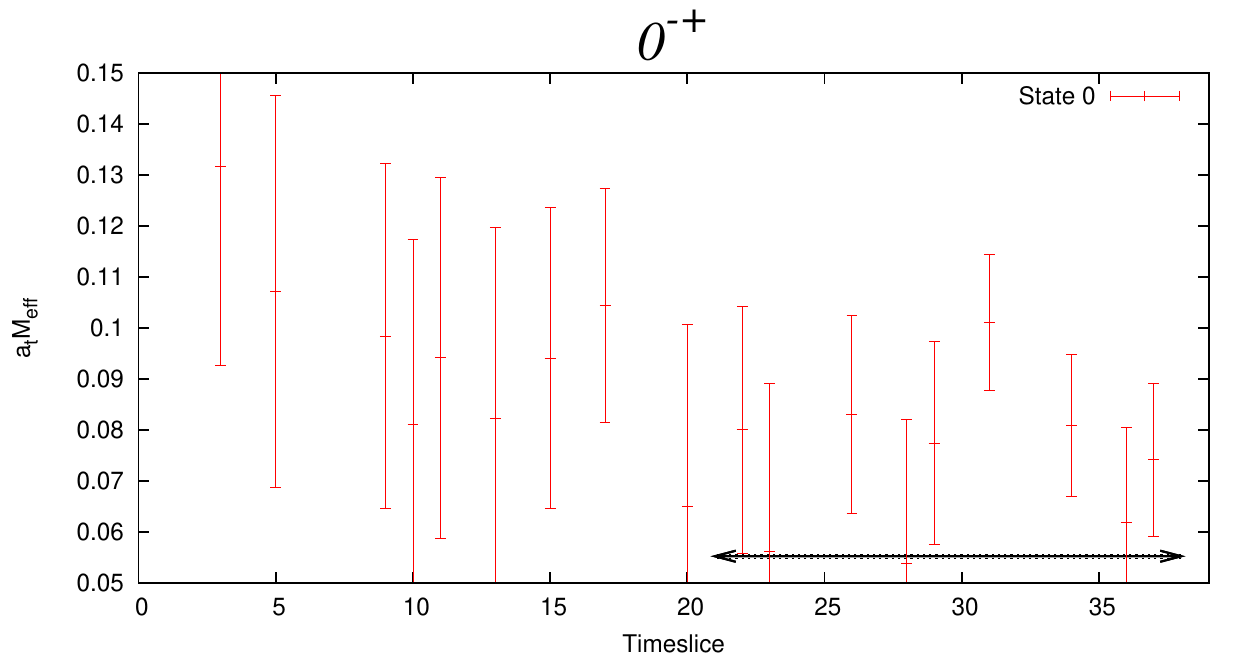} &
  \includegraphics[height=3.6cm,width= 6.9cm]{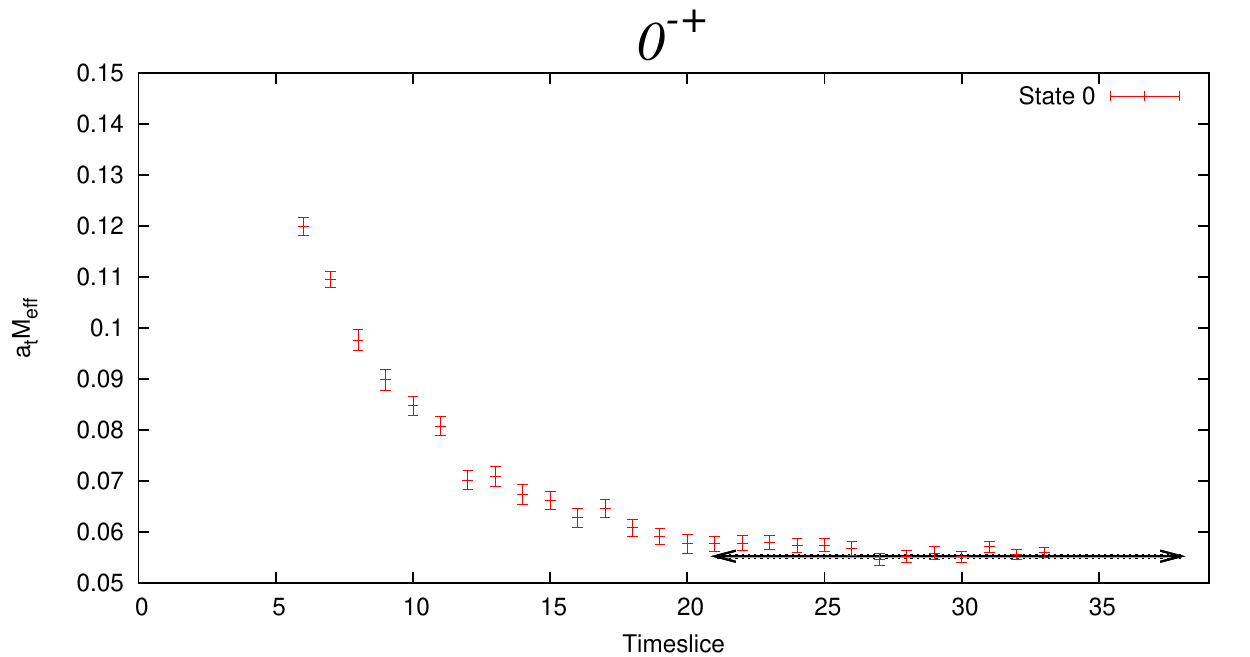}
  
\end{array}$
\caption{Enlarged version (left) of the effective mass plot for the pion ground state using the standard definition. On the right is an identical plot using the a 3-timeslice extended definition. The fit between timeslices 21 to 37 is shown by the black line with errors given by the dashed lines.}
\label{0mpb}
\end{figure*}

A sample correlation function, $C(t)$, for a pion on 96 configurations of $12^3\times 80$ dynamical anisotropic lattices can be seen in Fig.~\ref{0mp}  (these lattices are used in all results shown and the lattice parameters are detailed in Ref.~\cite{Juge:2005nr}). What is interesting to note however is that the use of time dilution has a dramatic  consequences for the effective mass plot shown on the right of this figure (defined here for a meson with periodic boundary conditions by $log(\frac{C(t+1)+C(t-1)}{2C(t)})$) . This is because we introduce independent stochastic noise at each timeslice. It is the correlation function that is fitted however and fitting between timeslices 21 to 37 yields the fit (with errors) in the effective mass plot on the left of Fig.~\ref{0mpb} . However, as can be seen on the right of Fig.~\ref{0mpb} , if we allow the effective mass plot to have more extended information by using next-to-nearest (or next-to-next-to-nearest) neighbours we can clean up the effective mass plot dramatically and our fit starts to look less outrageous. The reason for this is that the fit has information across an extended temporal region due to its fit window and can see through the temporal stochastic noise introduced by the stochastic estimator.

\section{A Fitting Robot}

The variational approach has a parameter space that can be explored by variation of the ``metric'' and ``optimising'' timeslices. Also the use of all-to-all propagators can dramatically extend the size of the operator  basis. Exploring these possibilities can lead to the creation of an extremely large number of correlation functions for fitting. At some point it becomes unfeasible to consider fitting these resultant correlation functions by  hand. In the particular case detailed above, the legitimacy of using the human eye and an effective mass plot to ascertain the accuracy of a fit that numerically gives a reasonable reduced $\chi^2$ value is immediately brought into question. It is for these reasons that we have investigated the creation of a fitting algorithm to fit correlation functions. The algorithm is based around 3 axioms, respected in the following order, 
\begin{itemize}\addtolength{\itemsep}{-0.8\baselineskip}
\item A reduced $\chi^2$ value below a given minimum signifies an acceptable fit,
\item The largest possible temporal window is to be chosen,
\item Plateaus beginning at early timeslices are preferred.
\end{itemize}

The first axiom is a standard fitting criterion applied to this type of data (here we use a value of 1.1 as the cut-off). The second axiom has a caveat in that only correlator data having a fractional error less than a chosen value (in the cases shown here 50\%) can be included in the fit. We do this because we must identify the limitations of our data; including terms that do not constrain the fit only serve to artificially lower the reduced $\chi^2$ value. This then chooses the maximum time value allowed in any fit region. The third axiom arises from the fact that the lower correlator values heavily constrain the fit (since the signal to noise ratio is decaying). If a large fit window exists that includes these points then they should be included to gain the full accuracy the data allows. Another reason for the third axiom is that the projected correlation matrix is not, in practice, completely diagonalised and it is possible that we can have contamination from lower-lying states. This can lead to a decay of the correlator into the lower state at large temporal values, dramatically effecting the accuracy of a fit.

In practice, the algorithm is implemented by first cutting the data at the appropriate temporal value (the last one with acceptable fractional error). Following our second axiom, we chose the largest possible temporal window and attempt a fit. Unless we have managed to exactly decompose the correlation matrix, such a fit will fail due to contamination from higher lying states. These states have greatest effect at low temporal values since they decay exponentially. Since such values are the most accurately determined (due to their signal to noise ratio), the inclusion of these low-lying points will always enlarge the reduced $\chi^2$ value because they are so heavily constraining. It should be noted that this in effect also allows us to chose the lowest time value allowed in the fit region to be as low as possible (but avoiding the contact term): any fit including early contaminated timeslices is destined to be rejected by virtue of the first axiom. Next we decrease the allowed fit window size and attempt all possible fit windows in the allowed region. We chose the ordering of these attempts such that the starting timeslice, $t_{min}$, of our fit window moves forward in time, i.e., we begin with $t_{min}$ at its lowest allowed value and then increase it until we have exhausted all possibilities. This then satisfies our third axiom. We iterate this procedure until we achieve an acceptable reduced $\chi^2$ value thus satisfying our first and most important axiom.

\subsection{Results}

In the results shown below, the operators employed are a combination of the point and extended operators detailed in Ref.~\cite{Juge:2005nr} and 4 different quark smearing levels (with a stout-link smeared gauge field). We use a quark mass in the region of the strange quark mass for all data. 

\begin{figure*}[ht]
\hspace{-1cm}
$\begin{array}{c@{\hspace{1cm}}c}
  \includegraphics[height=4.6cm,width= 7.9cm]{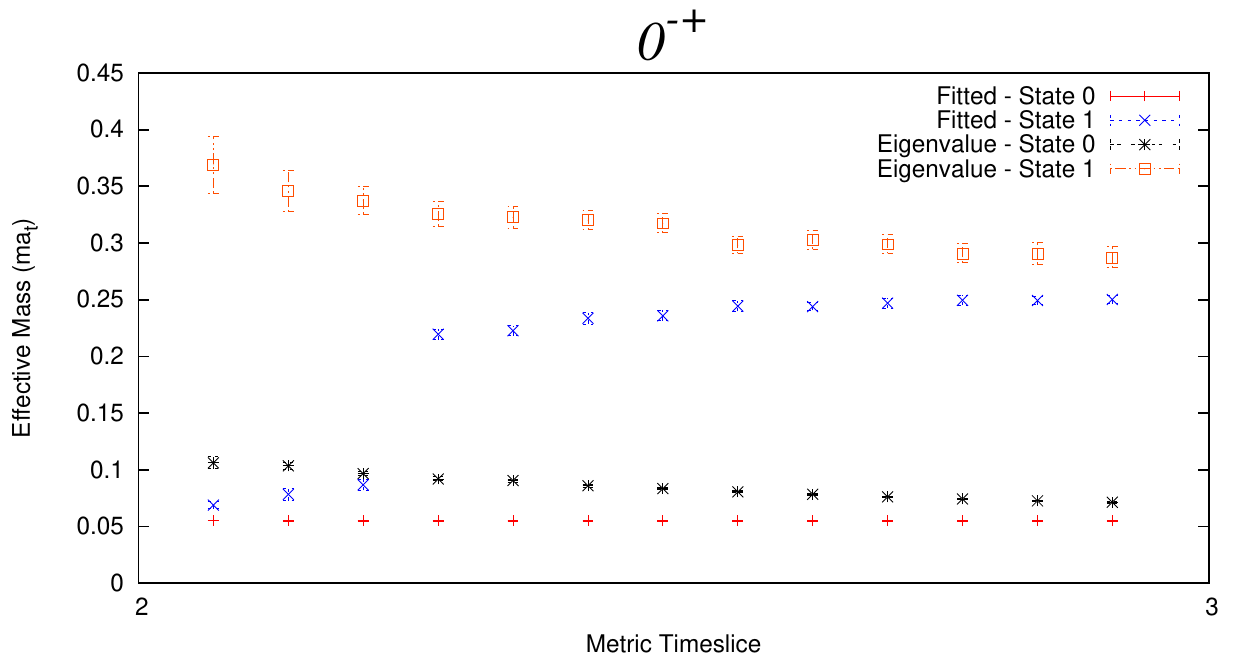} &
  \includegraphics[height=4.6cm,width= 7.9cm]{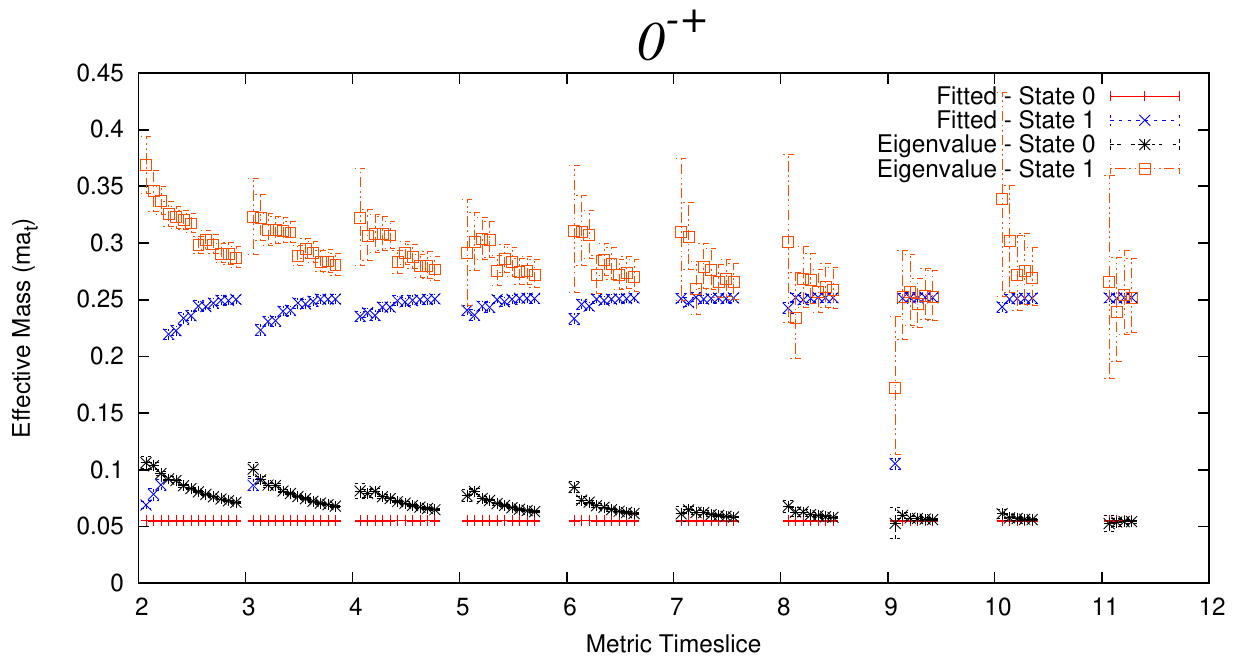}
  
\end{array}$
\caption{Plots of the fitted effective masses and eigenvalue masses as determined for a range of metric timeslices for the pion. The subdivisions correspond to different optimising timeslices.}
\label{VA0mp}
\end{figure*}

On the left of Fig.~\ref{VA0mp} we show the fitted effective masses of the pion, determined using the fitting robot from the projected correlation functions, for a single metric timeslice, 2, and a range of optimising timeslices. Along the X-axis we have the metric timeslices, $t_{met}$, with the sub-divisions between the points corresponding to different optimising timeslices, the first point being $(t_{met} + 1)$, the second $(t_{met} + 2)$, etc. We can directly observe the contamination of the eigenvectors by higher-lying states in the excited channel for low optimising timeslices. It is the plateauing from below of this effect as the optimising timeslice is increased that is interesting. For comparison we have also included the mass with error determined from the eigenvalues of the variational analysis. We can see that the eigenvalues are indeed approaching the same plateau regions.

On the right of Fig.~\ref{VA0mp} we compress the data of the left hand side and include further data from subsequent metric timeslices. For the ground state the fitted mass is consistent in all cases. The plateau effect in the excited channel is apparent at low metric timeslices but almost disappears from metric timeslice 7 onwards. We observe a consistency of these plateau regions across all metric timeslices. We also note that the eigenvalue effective masses begin to fall directly on the fitted regions as the metric timeslice increases. While the errors on these eigenvalues increase, the errors determined from the projected data remain unaffected.

\begin{figure*}[ht]
\hspace{-1cm}
$\begin{array}{c@{\hspace{1cm}}c}
  \includegraphics[height=4.6cm,width= 7.9cm]{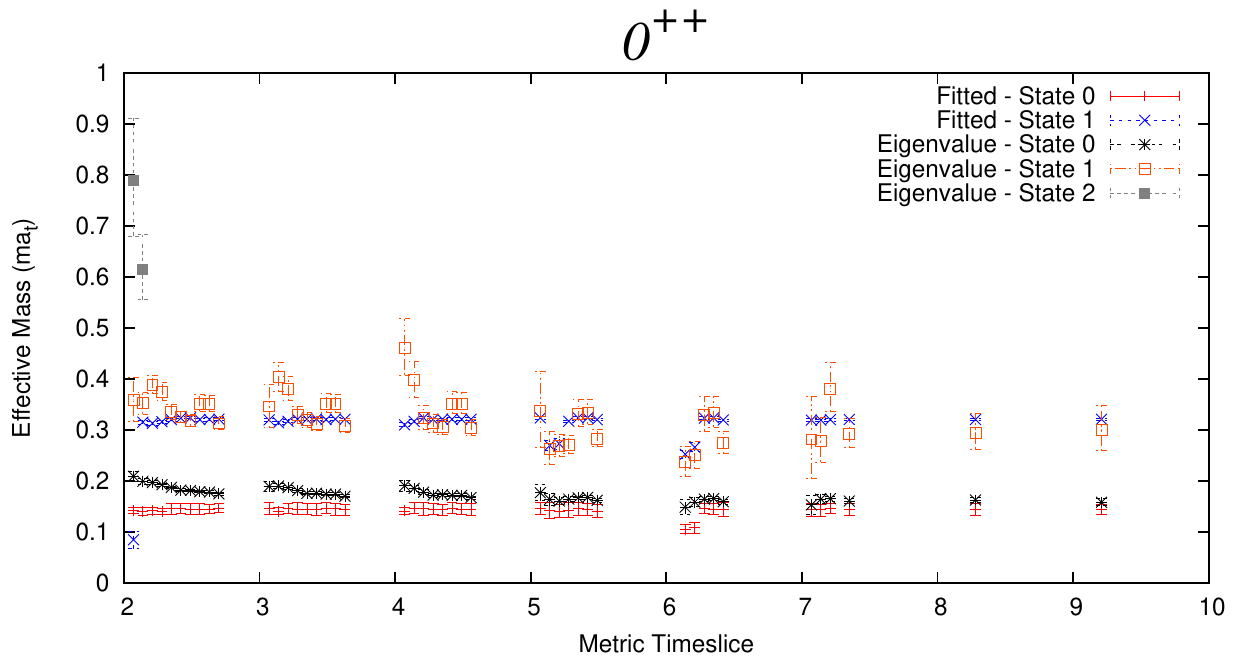} &
  \includegraphics[height=4.6cm,width= 7.9cm]{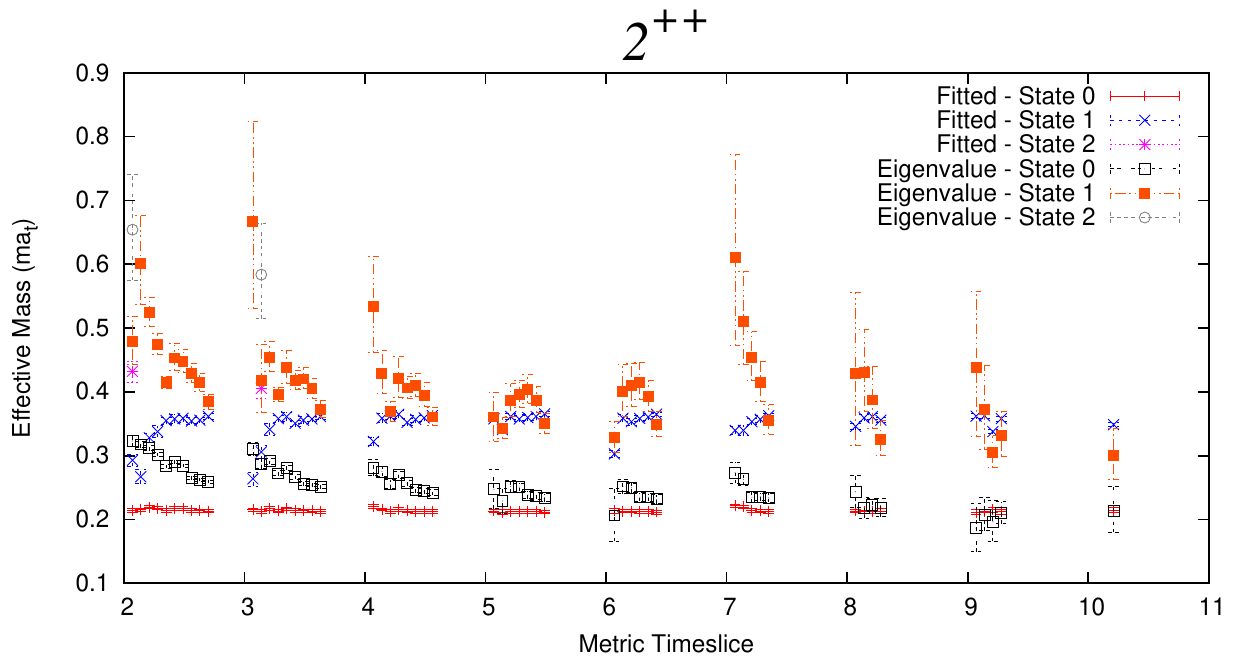}
  
\end{array}$
\caption{As before, for the $0^{++}$ (left) and $2^{++}$ (right) states.}
\label{VAops}
\end{figure*}

These operators above were just different smearing combinations of the standard $\gamma_5$ pion operator since this is known to have a high signal-to-noise ratio. We have also carried out similar analysis for a range of states and having included extended operators. Here we show some of the more interesting results. In Fig.~\ref{VAops} we can see similar behaviour to the pion for the $0^{++}$ isovector state and the $2^{++}$ $P$-wave state. Only slight variations in the overall behaviour occur in both cases. For the $0^{++}$, we see that the effective masses determined from the eigenvalues drops in a few cases at timeslices 5 and 6. At metric 6 this is accompanied by a disturbing drop in the determined ground state energy. However, since the overall trend is obvious, we can still accurately chose fit values for the ground and first-excited states. In the case of the $P$-wave we observe slightly more erratic behaviour of the fit to the projected data, coupled again with erratic behaviour of the eigenvalues. In this particular case, three operators were used in the correlation matrix but after the first few very early successes the robot was unable to find suitable fits of the projected data. Resulting errors from poor diagonalisations may have caused the effects that we observe in this case.

\begin{figure*}[ht]
\hspace{-1cm}
$\begin{array}{c@{\hspace{1cm}}c}
  \includegraphics[height=4.6cm,width= 7.9cm]{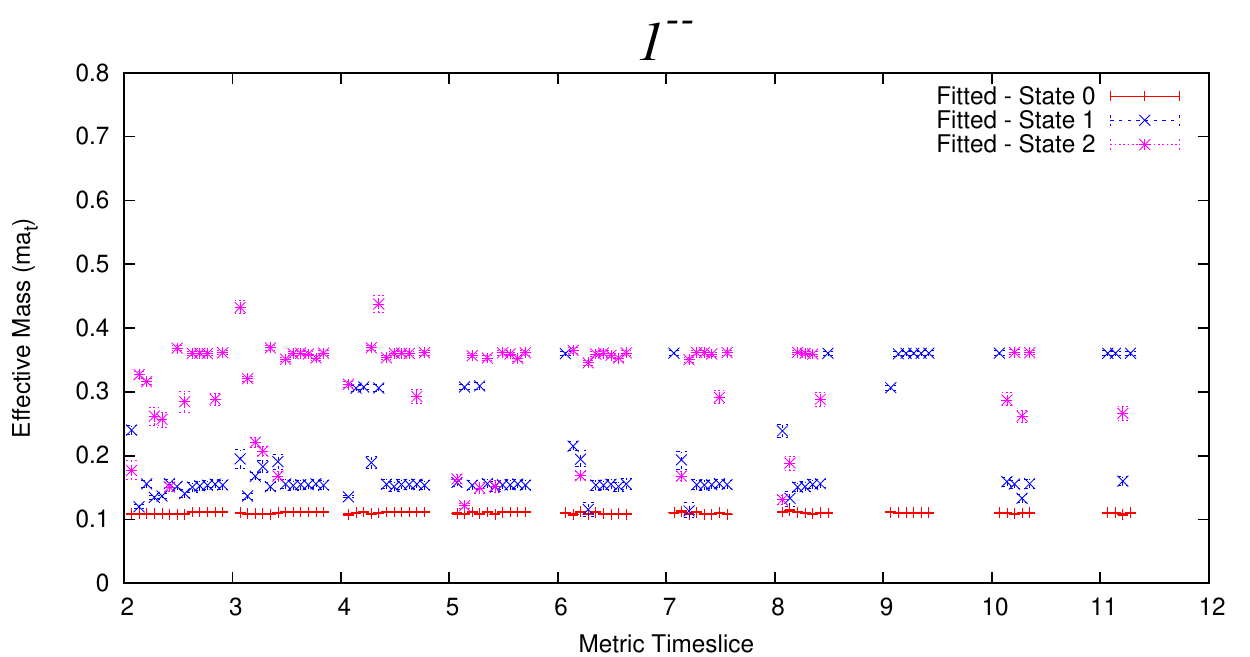} &
  \includegraphics[height=4.6cm,width= 7.9cm]{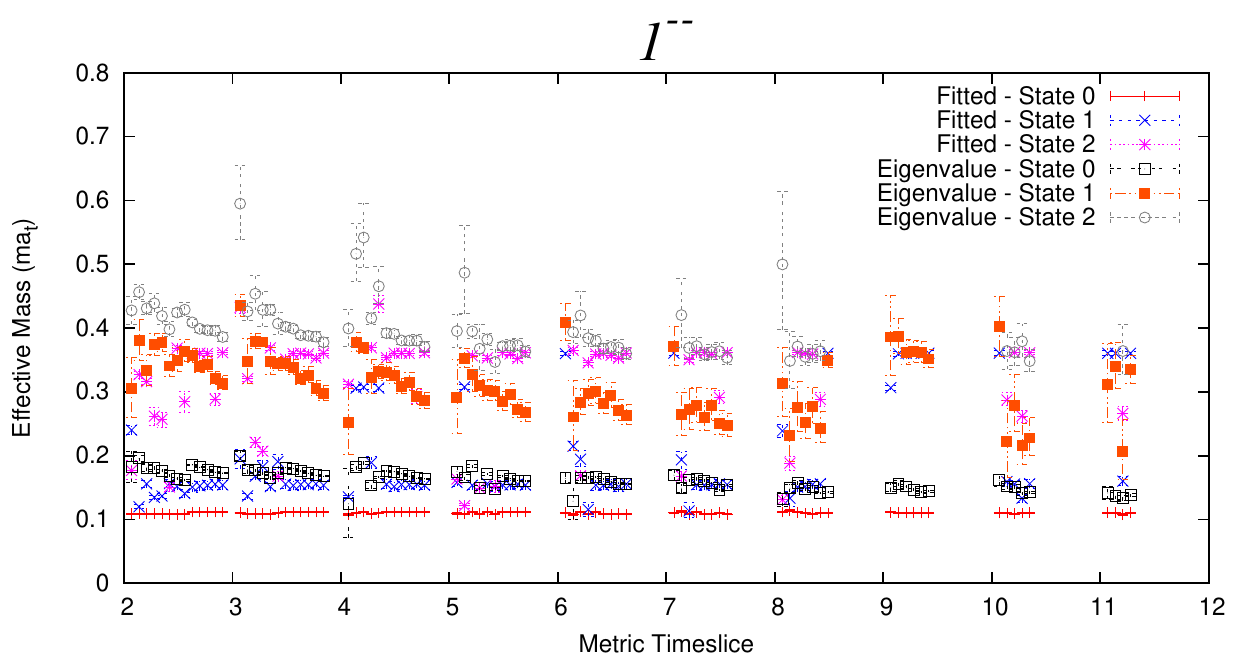}
  
\end{array}$
\caption{As before, the figure on the left is plotted without eigenvalue masses.}
\label{VA1mm}
\end{figure*}

Lastly, in Fig.~\ref{VA1mm} we also include results for the $\rho$ meson where we have large data sets for 3 operators. On the left we have the fitted mass values; on the right we have the same data but with the eigenvalue masses overlaid. In the cases where we were unsuccessful in the diagonalisation of the $3\times3$ correlation matrix we have reduced it to a 2 operator diagonalisation (for example all of metric 9). On the left, again we see some erratic behaviour in plenty of cases but at mid range metrics and higher optimising timeslices we see quite systematic behaviour and accurate determination of 3 states. Interestingly, on the right graph, it appears that the lowest eigenvalue is plateauing at the first-excited state fitted mass (even in the case of metric 9), and that none of the eigenvalues approach the ground state value that is determined so consistently across all cases. 

\section{Conclusions}

The fitting robot functions consistently in our analysis. With any such attempt however, it is inevitable that numerical situations will arise that circumvent all axioms of the robot due to a spuriously low reduced $\chi^2$ value. It is for this reason that the method couples so well with the variational approach. This allows for multiple fittings of data derived from an identical source and we can then observe plots such as those in Fig.~\ref{VAops} and use the human eye as the ultimate test to determine where the fit truly lies. Improvements to the robot should include a consistency check of the chosen window - to identify incorrect fits a bootstrap of the chosen data points should be performed to search for excessive chi-squared and fitted mass variations. To assist with this, a fit to the principal correlators for each metric timeslice should also be included on the derived figures.

This work has yielded some very interesting insights into the variational approach. These results indicate that the robustness of the choice of variational parameters should always be checked. It also complements Refs.~\cite{Blossier:2008tx, Dudek:2007wv} in their discussions and conclusions with respect to the selection of the metric timeslice. We believe that with the suggested improvements to the fitting robot and a larger and more probing selection of the operator basis that such an approach can yield consistent and robust results from the analysis of correlation matrices.


\begin{thebibliography}{99}\setlength{\itemsep}{-2mm}



\bibitem{Luscher:1990ck}
  M.~Luscher and U.~Wolff,
  Nucl.\ Phys.\  B {\bf 339}, 222 (1990).

\bibitem{Blossier:2008tx}
  B.~Blossier, G.~von Hippel, T.~Mendes, R.~Sommer and M.~Della Morte,
  arXiv:0808.1017 [hep-lat].

\bibitem{Wilcox:1999ab}
  W.~Wilcox,
  arXiv:hep-lat/9911013.

\bibitem{Neff:2001zr}
  H.~Neff, N.~Eicker, T.~Lippert, J.~W.~Negele and K.~Schilling,
  Phys.\ Rev.\  D {\bf 64}, 114509 (2001)
  [arXiv:hep-lat/0106016].

\bibitem{Foley:2005ac}
  J.~Foley, K.~Jimmy Juge, A.~O'Cais, M.~Peardon, S.~M.~Ryan and J.~I.~Skullerud,
  Comput.\ Phys.\ Commun.\  {\bf 172}, 145 (2005)
  [arXiv:hep-lat/0505023].

\bibitem{Collins:2007mh}
  S.~Collins, G.~Bali and A.~Schafer,
  PoS {\bf LAT2007} (2007) 141
  [arXiv:0709.3217 [hep-lat]].

\bibitem{Boyle:2008rh}
  P.~A.~Boyle, A.~Juttner, C.~Kelly and R.~D.~Kenway,
  JHEP {\bf 0808} (2008) 086
  [arXiv:0804.1501 [hep-lat]].

\bibitem{Juge:2005nr}
  K.~J.~Juge, A.~O'Cais, M.~B.~Oktay, M.~J.~Peardon and S.~M.~Ryan,
  PoS {\bf LAT2005}, 029 (2006)
  [arXiv:hep-lat/0510060].

\bibitem{Dudek:2007wv}
  J.~J.~Dudek, R.~G.~Edwards, N.~Mathur and D.~G.~Richards,
  Phys.\ Rev.\  D {\bf 77} (2008) 034501
  [arXiv:0707.4162 [hep-lat]].

\end{thebibliography}
\end{document}